\documentclass[a4paper,10pt,twoside]{cpc-hepnp}

\usepackage{multicol}
\usepackage{graphicx}
\usepackage{booktabs}
\usepackage{amssymb,bm,mathrsfs,bbm,amscd,multirow}
\usepackage[tbtags]{amsmath}
\usepackage{lastpage}
\newcommand{\Nsoneabf}{$N_{1/2^-}(1535)$}
\newcommand{\Nsonebbf}{$N_{1/2^-}(1650)$}

\newcommand{\bc}           {\begin{center}}
\newcommand{\ec}           {\end{center}}
\newcommand{\bq}           {\begin{eqnarray}}
\newcommand{\eq}           {\end{eqnarray}}
\newcommand{\be}           {\begin{equation}}
\newcommand{\ee}           {\end{equation}}
\newcommand{\bi}           {\begin{itemize}}
\newcommand{\ei}           {\end{itemize}}

\newcommand{\Nponecbf}{$N_{1/2^+}(1880)$}

\newcommand{\Npthreeabf}{$N_{3/2^+}(1720)$}
\newcommand{\Npthreebbf}{$N_{3/2^+}(1900)$}

\newcommand{\Ndthreeabf}{$N_{3/2^-}(1520)$}
\newcommand{\Ndthreebbf}{$N_{3/2^-}(1700)$}

\newcommand{\Nffiveabf}{$N_{5/2^+}(1680)$}
\newcommand{\Nffivebbf}{$N_{5/2^+}(2000)$}

\newcommand{\Ndfiveabf}{$N_{5/2^-}(1675)$}

\newcommand{\Nfsevenabf}{$N_{7/2^+}(1990)$}

\newcommand{\Dsoneabf}{$\Delta_{1/2^-}(1620)$}
\newcommand{\Dsonebbf}{$\Delta_{1/2^-}(1900)$}

\newcommand{\Dponebbf}{$\Delta_{1/2^+}(1910)$}

\newcommand{\Dpthreecbf}{$\Delta_{3/2^+}(1920)$}
\newcommand{\Ddthreeabf}{$\Delta_{3/2^-}(1700)$}
\newcommand{\Ddthreebbf}{$\Delta_{3/2^-}(1940)$}

\newcommand{\Dffiveabf}{$\Delta_{5/2^+}(1905)$}

\newcommand{\Ddfiveabf}{$\Delta_{5/2^-}(1930)$}

\newcommand{\Dfsevenabf}{$\Delta_{7/2^+}(1950)$}

\newcommand{\Dgsevenabf}{$\Delta_{7/2^-}(2200)$}

\begin{document}

\fancyhead[co]{\footnotesize E. Klempt: Nucleon Excitations}

\footnotetext[0]{Received \today}

\title{Nucleon excitations\thanks{Supported within SFB/TR16 by the Deutsche
Forschungsgemeinschaft}}

\author{%
      Eberhard Klempt\email{klempt@hiskp.uni-bonn.de}%
} \maketitle

\address{%
Helmholtz-Institut f\"ur Strahlen- und Kernphysik der Universit\"at
Bonn\\
Nu\ss allee 14-16, 53115 Bonn, Germany
\\
}

\begin{abstract}
The mass pattern of nucleon and $\Delta$ resonances is compared with
predictions based on quark models, the Skyrme model, AdS/QCD, and
the conjecture of chiral symmetry restoration.
\end{abstract}



\begin{multicols}{2}

\section{Introduction}
The driving force behind the intense efforts to clarify the spectrum
of meson and baryon resonances is the aim to improve the
understanding of the confinement mechanism and of the dynamics of
quarks and gluons in the non-perturbative region of QCD. Different
approaches have been developed.

The systematics of the baryon ground states were constitutive for
the development of quark models. For excited states, different quark
model variants are capable of reproducing the main features of the
data but the models fail in important details: the number of
expected states is considerably larger than confirmed
experimentally, and the masses of radial excitations are mostly
predicted at too high masses.

Resonances fall into a mass range where the usefulness of quarks and
gluons can be debated; there are attempts to generate resonances
dynamically from ground-state mesons (pseudoscalar and vector) and
ground-state baryons (octet and decuplet). Possibly, this is an
alternative approach to the resonance spectrum; the mechanism may
however also be the source of additional resonances which come atop
of the quark model states.

In the harmonic oscillator (h.o.) approximation, the quark model
predicts a ladder of meson and baryon resonances with equidistant
squared masses, alternating with positive and negative parity, and
this pattern survives in more realistic potentials. Experimentally,
positive and negative parity states are often degenerate in mass.
This fact is the basis for the conjecture that chiral symmetry might
be restored when resonances are excited into the high-mass region.

AdS/CFT is a new approach to describe QCD phenomena in an
analytically solvable model over a wide range of interaction
energies. The calculations include the meson and baryon mass
spectrum. In the case of mesons, most masses (except those for
scalar and pseudoscalar mesons) are well reproduced; for baryons the
numerical success is amazing.

Finally, there was the claim at this conference that the Skyrme
model does as well as  AdS/QCD in reproducing the mass spectrum.
Hence the comparison, predictions versus experiment, will be done
for predictions of AdS/QCD, of the Skyrme model, and of three
different quark model variants. The main focus of the talk will be
on baryon resonances; mesons will be mentioned briefly in a few
cases.

\section{Quark models}

In the quark model, there are two independent oscillators. Choosing
harmonic oscillators, the states are characterized by
$$
(D,L^P_{\textsf{N}})
$$
where $D$ is the SU(3) dimensionality (56 or 70), $L$ the orbital
angular momentum, $P$ the parity, $\textsf{N}$ the shell number. Two
questions emerge. First, can we relate these h.o. states with
observed resonances? Second, is there some systematic of the
so-called missing resonances? We remind the reader that not all
solutions of a Hamiltonian need to be realized dynamically. The
observed states {\it and} the missing states may hence contain an
important message how QCD arranges a three-quark system at large
excitation energies.

There are great successes of the quark model for baryon
spectroscopy: these include the interpretation of ground-state
baryons in SU(3) multiplets, the correct prediction of the
multiplicity of low-mass negative-parity states in the first
excitation band $(D,L^P_{\textsf{N}})=(70,1^-_1)$, and the correct
prediction of baryon properties like formfactors, magnetic moments.
But some problems remain\vspace{-2mm} \bi \item The \Nsoneabf\ --
$N_{1/2^+}(1440)$\footnote{\footnotesize{We give spin and parity of
a resonance explicitly and not the $N\pi$ partial wave.}} mass
difference is +100\,MeV experimentally, and -\,80\,MeV in most quark
models.\vspace{-2mm}
\item There are more states predicted than found experimentally
(missing resonance problem).\vspace{-2mm}
\item There are no states in a 20-plet, expected from the SU(6)
decomposition  $6\otimes6\otimes6=56\oplus2\cdot 70\oplus
20$.\vspace{-2mm}
\item Conceptually, one may ask if constituent quarks should have a
defined rest mass when going to high excitation
energies.\vspace{-2mm}
\end{itemize}

Quark model variants help to improve some details. The gluonic flux
tube can be excited leading to a rich spectrum of hybrid baryons but
this possibility aggravates the problem of the missing resonances.
Five-quark components in the wave functions can be justified since a
$P$-wave excitation in $(q-qq)$ ``costs" about 450 MeV, and adding a
pseudoscalar $q\bar q$ pair in $S$-wave in $(q\bar qqqq)$ may be
energetically favored. Baryons like \Nsoneabf\ could thus be made up
from five quarks. If these cluster into a $qqq$ and a $q\bar q$
color singlet, these states are not necessarily due to multi-quark
chemistry but rather meson-nucleon molecules.

\section{Dynamically generated resonances}
The classical example for a dynamically generated resonance is the
$\Delta(1232)$ which is represented as $qqq$ state in quark models
while Chew explained it as resonance in the $N\pi$
system\cite{Chew:1955zz}. In the modern concept, nucleon and
$\Delta(1232)$ are considered as fundamental particles from which
higher-mass resonances are constructed. An often discussed state is
$N_{1/2^-}(1535)$ which can be very successfully described as
$N\eta$-$\Sigma K$ coupled-channel effect\cite{Kaiser:1995eg} or
$\Lambda_{1/2^-}(1405)$ coupling strongly to $\Sigma\pi$ and $N\,K$.
Possibly, the latter resonance is split into two
states\cite{Jido:2003cb}. Open is the question if all baryon
resonances can be constructed from their decay modes. And it is also
unclear if the generation of resonances provides a dual description
of the same baryons as the quark model or if $qqq$ and molecular
descriptions lead to different states which could co-exist (and may
mix) leading to a larger number of states than predicted by quark
models alone.

Here it must be mentioned that quark model states need long-range
corrections with higher Fock configurations. These are dominated by
the meson--baryon interaction (and include four-quark and hybrid
configurations). Resonances described in a hadronic picture require
short-range corrections. These lead back to interacting quarks and
gluons. Hence quark-model wave functions and meson--baryon states
have a sizable overlap and possibly, they span the same Hilbert
space. Finally, both chiral Lagrangians and quark-model Lagrangians
are approximations of the same underlying theory, of QCD. The
resulting spectra should not be just added.

This view has far reaching consequences. It is easy to accept that
there is one $N_{1/2^-}(1535)$ resonance, not a quark model state
and a dynamically generated one. If $f_0(980)$ and $a_0(980)$, and
the hidden charm resonances $X,Y,Z$, are both molecules and $q\bar
q$, this is a controversially debated question. $\sigma(500)$ and
$\kappa(700)$ are certainly dynamically generated and not $1^3P_0$
quark model states. A different question is what happens when the
current quark mass could be changed continuously from the $b$-quark
mass to light quarks. It is possible that this gedanken experiment
would connect the $\chi_{b0}(1P)$ to the $\sigma$. The response of
QCD certainly depends critically on the mass when a $q\bar q$ pair
is created in the vacuum. In the light quark sector, $\sigma$ is the
lowest mass state, in the bottomonium sector it is $\chi_{b0}(1P)$,
hence $\sigma$ may deserve the notation and interpretation as
$f_{0}(1P)$ state.

\section{Chiral multiplets}
Baryon resonances exhibit an unexpected phenomenon: parity doublets,
pairs of resonances with the same spin $J$ but opposite
parities\cite{Glozman:this}. Often, these are quartets of $N^*$ and
$\Delta^*$ having the same $J$, see Table \ref{tab1}. Resonances and
star rating are taken from PDG\cite{Amsler:2008zzb}. If only those
states are included which survived the latest GWU
analysis\cite{Arndt:2006bf}, no quartet and only few parity doublets
remain. The chiral multiplets are interpreted as indication that
chiral symmetry may be restored at large excitation energy.

\bc\tabcaption{\label{tab1}Chiral multiplets for $J=1/2$, $3/2$, $5/2$ (first three lines) and for
$J=1/2,\cdots , 7/2$ (last four lines) for nucleon and $\Delta$
resonances. }
\begin{footnotesize}
\renewcommand{\arraystretch}{1.1}
\begin{tabular}{cccc} \toprule
\hspace{-2mm}N$_{1/2^+}(1710)$&\hspace{-3mm}N$_{1/2^-}(1650)$&\hspace{-3mm}$\Delta_{1/2^+}(1750)$&\hspace{-3mm}$\Delta_{1/2^-}(1620)$\vspace{-1.5mm}\hspace{-2mm}\\
\hspace{-2mm}**&****&&****\vspace{-1mm}\hspace{-2mm}\\
\hspace{-2mm}N$_{3/2^+}(1720)$&\hspace{-3mm}N$_{3/2^-}(1700)$&\hspace{-3mm}$\Delta_{3/2^+}(1600)$&\hspace{-3mm}$\Delta_{3/2^-}(1700)$\vspace{-1.5mm}\hspace{-2mm}\\
\hspace{-2mm}****&***&***&****\vspace{-1mm}\hspace{-2mm}\\
\hspace{-2mm}N$_{5/2^+}(1680)$&\hspace{-3mm}N$_{5/2^-}(1675)$&\multicolumn{2}{c}{no chiral partners}\vspace{-1.5mm}\hspace{-2mm}\\
\hspace{-2mm}****&****&&\vspace{-1mm}\hspace{-2mm}\\
\hspace{-2mm}N$_{1/2^+}(1880)$&\hspace{-3mm}N$_{1/2^-}(1905)$&\hspace{-3mm}$\Delta_{1/2^+}(1910)$&\hspace{-3mm}$\Delta_{1/2^-}(1900)$\vspace{-1.5mm}\hspace{-2mm}\\
\hspace{-2mm}**&*&****&**\vspace{-1mm}\hspace{-2mm}\\
\hspace{-2mm}N$_{3/2^+}(1900)$&\hspace{-3mm}N$_{3/2^-}(1860)$&\hspace{-3mm}$\Delta_{3/2^+}(1920)$&\hspace{-3mm}$\Delta_{3/2^-}(1940)$\vspace{-1.5mm}\hspace{-2mm}\\
\hspace{-2mm}**&**&***&**\vspace{-1mm}\hspace{-2mm}\\
\multicolumn{2}{c}{no chiral partners}&\hspace{-3mm}$\Delta_{5/2^+}(1905)$&\hspace{-3mm}$\Delta_{5/2^-}(1930)$\vspace{-0.5mm}\hspace{-2mm}\\
\hspace{-2mm}&&****&***\vspace{-1mm}\hspace{-2mm}\\
\hspace{-2mm}N$_{7/2^+}(1990)^a$&\hspace{-3mm}N$_{7/2^-}(2190)$&\hspace{-3mm}$\Delta_{7/2^+}(1950)$&\hspace{-3mm}$\Delta_{7/2^-}(2200)$\vspace{-1.5mm}\hspace{-2mm}\\
\hspace{-2mm}**&****&****&*\vspace{-1mm}\hspace{-2mm}\\
\hspace{-2mm}N$_{9/2^+}(2220)$&\hspace{-3mm}N$_{9/2^-}(2250)$&\hspace{-3mm}$\Delta_{9/2^+}(2300)$&\hspace{-3mm}$\Delta_{9/2^-}(2400)$\vspace{-1.5mm}\hspace{-2mm}\\
\hspace{-2mm}****&****&**&**\vspace{-1mm}\hspace{-2mm}\\
\bottomrule
\end{tabular}
\renewcommand{\arraystretch}{1.0}
\end{footnotesize}
\ec

\section{Super-multiplets with defined
quantum numbers}
\subsection{$\vec L$ and $\vec S$}
Relativity plays an important role in quark models. In relativistic
models, only the total angular momentum $J$ is defined.
Experimentally, there are a few striking examples where the leading
orbital angular momentum and the spin can be identified (small
admixtures of other components are not excluded).\\ 1. The
negative-parity light-quark baryons, collected in the first data
block of Table \ref{tab2}, form a $N^*$ doublet, a $N^*$ triplet,
and a $\Delta^*$ doublet, well separated in mass from all
other negative parity states.\\
2. The positive parity states (second block) form an isolated $N^*$
doublet, a
$N^*$ quartet, and a $\Delta^*$ quartet.\\
3. At higher mass there is a mass degenerate negative-parity
$\Delta^*$ triplet and a $\Delta^*$ doublet (third block).\\
These multiplets are separated by 200 MeV from other states having
the same quantum numbers. Of course, mixing of states having
identical quantum numbers is possible; but there is no visible
effect of mixing on the masses.

Frequently a statement is made that $L$ and $S$ cannot be good
quantum numbers. Quarks, even constituent quarks, are supposed to
move with relativistic velocities. And in relativity, only $J$ is
defined. But we should admit that we do not know the dynamical
origin of the mass of a resonance. The nucleon mass is predominantly
due to field energy. Why should the mass of the \Dfsevenabf\ not be
predominantly due to field energy? As long as we have no deep
understanding of the mechanism leading to the excited states, we
should take phenomenology serious. And phenomenologically, $L,S$
supermultiplets are an important organizing principle for baryon
spectroscopy.

\subsection{The radial
excitation quantum number $N$} In the harmonic oscillator
approximation, a shell number \textsf{N} is defined which gives - to
first order - the masses of baryon resonances. We use, instead, the
\end{multicols}

\begin{center}
\tabcaption{ \label{tab2}Supermultiplets in $L$ and $S$ for N and
$\Delta$ excitations (upper part). The lower part of the table shows
the mass square splitting of states within a given partial wave.}
\renewcommand{\arraystretch}{1.15}
\footnotesize
\begin{tabular*}{170mm}{@{\extracolsep{\fill}}cccccccc}
\toprule
& $L; S$     & $J^P=1/2^-$ &$J=3/2^-$ &$J=5/2^-$ \vspace{1mm}\\
\hline
&$L=1; S=1/2$&\Nsoneabf & \Ndthreeabf &  \\
&$L=1; S=3/2$& \Nsonebbf & \Ndthreebbf & \Ndfiveabf\\
&$L=1; S=1/2$&\Dsoneabf & \Ddthreeabf& \\
\hline
&$L; S$     & $J^P=1/2^+$ &$3/2^+$ &$5/2^+$&$7/2^+$ \\
\hline
&$L=2; S=1/2$&&\Npthreeabf & \Nffiveabf &  \\
&$L=2; S=3/2$&\Nponecbf& \Npthreebbf &\Nffivebbf &\Nfsevenabf\\
&$L=2; S=3/2$&\Dponebbf & \Dpthreecbf & \Dffiveabf& \Dfsevenabf\\
\hline
&&$J^P=1/2{^-}$&$3/2{^-}$& $5/2^-$&$7/2^-$\\
\hline
&$L=1; S=3/2$&\Dsonebbf&\Ddthreebbf&\Ddfiveabf&\vspace{-1mm}No state!\\
&$L=3; S=1/2$&&&$\Delta_{5/2^-}(2233)$&\Dgsevenabf\vspace{1mm}\\
\toprule
&&$N, \Delta$&$\Lambda$&$\Sigma, \Sigma^*$&$\Xi, \Xi^*$&N=0\vspace{1mm}\\
\hline &56, 8; 1/2&$N_{1/2^+}(1440)$&$\Lambda_{1/2^+}(1600)$&
$\Sigma_{1/2^+}(1660)$ & $\Xi_{1/2^+}(1690)$\vspace{-1mm}
&\multirow{4}{*}{N=1} \\
&\scriptsize$\delta M^2$&\scriptsize 1.19$\pm$0.11&\scriptsize
1.31$\pm$0.11& \scriptsize1.34$\pm$0.11&\scriptsize 1.13$\pm$0.03 \\
&56, 10; 3/2&$\Delta_{3/2^+}(1600)$&&$\Sigma_{3/2^+}(1840)$&x\\
&\scriptsize$\delta M^2$&\scriptsize$1.04\pm0.15$&&\scriptsize$1.47\pm0.44$&\\
\hline & 70, 8; 1/2&$ N_{1/2^+}(1710)$&$\Lambda_{1/2^+}(1810)$&
$\Sigma_{1/2^+}(1770)$ & x\vspace{-1mm}
&\multirow{2}{*}{Possibly}\hspace{-50mm}\\
&\scriptsize$\delta M^2$&\scriptsize 2.04$\pm$0.15&\scriptsize
2.03$\pm$ 0.15&\scriptsize $1.72\pm0.16$&&\multirow{2}{*}{ N=2}\hspace{-50mm}\\
&70, 10; 1/2&$\Delta_{1/2^+}(1750)$&&$\Sigma_{1/2^+}(1880)$&x\vspace{-1mm}\\
&\scriptsize$\delta
M^2$&\scriptsize$1.54\pm0.16$&&\scriptsize2.12$\pm$0.11\\\bottomrule
\end{tabular*}
\renewcommand{\arraystretch}{1.0}
\ec

\begin{multicols}{2}
\noindent radial excitation number $N$. $N$ and \textsf{N} are
related by $\textsf{N}= L + 2 N$. To make contact with models, we
define $N=0$ for the lowest-mass state. The Roper-like resonances
(lowest mass states with ground-state q.n.: $N_{1/2^+}(1440)$,
$\Delta_{3/2^+}(1600)$, $\Lambda_{1/2^+}(1600)$,
$\Sigma_{1/2^+}(1660)$, $\Xi_{1/2^+}(1690)$) are given in the forth
data block in Table \ref{tab2}. The spacings are all compatible with
the spacing, per unit of angular momentum, of the leading (meson or)
baryon trajectory (which is 1.14\,GeV$^2$).

The last (fifth) data block gives the third state in a given partial
wave. For the second radial excitation, the expected spacing w.r.t.
the ground state would be 2.28\,GeV$^2$. In the quark model, the
states would belong to the fifth excitation band and the expected
spacing would be in the order of 5.5\,GeV$^2$. The quark model
suggests, however, states in which the two intrinsic harmonic
oscillators are orbitally excited to $l_1=l_2=1$ and that $\vec
L=\vec l_1+\vec l_2$ vanishes. In this case, the states belong to a
70-plet in SU(6). In this interpretation, also two states with $\vec
L=\vec l_1+\vec l_2$ and $L=1$, i.e. with $J=1/2^+$ and $J=3/2^+$
should be observed. Since $L=2$ gives $\approx$1930\,MeV, $L=0$
$\approx$1730\,MeV, we may expect such a doublet at about 1830\,MeV.
Since both oscillators are excited, they may decouple from
single-pion emission and could be observable in a cascade only, e.g.
via \Ndthreeabf$\pi$.

\subsection{Can all these data be used?}

The recent analysis of the GWU group has shed doubts on the
existence of many of the states reported in the Karlsruhe-Helsinki
and Carnegie Mellon analyses \cite{Hohler:1979yr,Cutkosky:1980rh}.
Of course, it is an open question if the old analyses are right or
if many states listed in the Review of Particle Properties
\cite{Amsler:2008zzb} are fake. In the BnGa partial wave analysis
\cite{Anisovich:2009pr} many resonances, not seen in the GWU
analysis, do show up in inelastic reactions. For the time being, the
evidence for a failure of the old analysis is not convincing, and
the full spectrum listed in \cite{Amsler:2008zzb} is used for the
discussion presented here.

\section{AdS/QCD}
The AdS/CFT correspondence provides an analytically solvable
approximation to QCD in the regime where the QCD coupling is large.
It has led to important insights into the properties of quantum
chromodynamics and can be used to calculate the hadronic spectrum of
light-quark meson and baryon resonances\cite{Brodsky:this}. The
dynamics is controlled by a variable $\zeta$ which is
suggested\cite{Brodsky:this} to be related to the mean distance
between the constituents.  In the hard-wall approximation, $\zeta$
is constrained to $\zeta\leq\zeta_{\rm max}=1/\Lambda_{\rm QCD}$. In
the soft wall approximation\cite{Karch:2006pv}, a dilaton background
field proportional to $\zeta^2$ is introduced which limits the mean
distance between the constituents softly. The results on the baryon
excitation spectrum shown below refer to solutions with a soft wall.

\subsection{$\Delta$ resonances}
Applied to $\Delta$ resonances, a very simple formula can be
derived\cite{Forkel:2007cm} which reads \begin{equation} M^2 =
1.04\cdot ({L}+{N}+3/2)\,\left[{\mathrm GeV^2}\right]\,.
\label{eq1}
\end{equation}
Replacing 3/2 by 1/2 and with a small readjustment of numerical
constant by less than 10\%\cite{Forkel:2007cm}, the meson mass
spectrum is reproduced qualitatively, except for scalar and
pseudoscalar mesons (see Fig. 57 in ref.\cite{Klempt:2007cp}). For
the $\Delta$ excitation spectrum, the agreement is excellent as
visualized in Fig. \ref{fig1}. To $\Delta$ reso-
\end{multicols}

\begin{center}
\includegraphics[width=12cm]{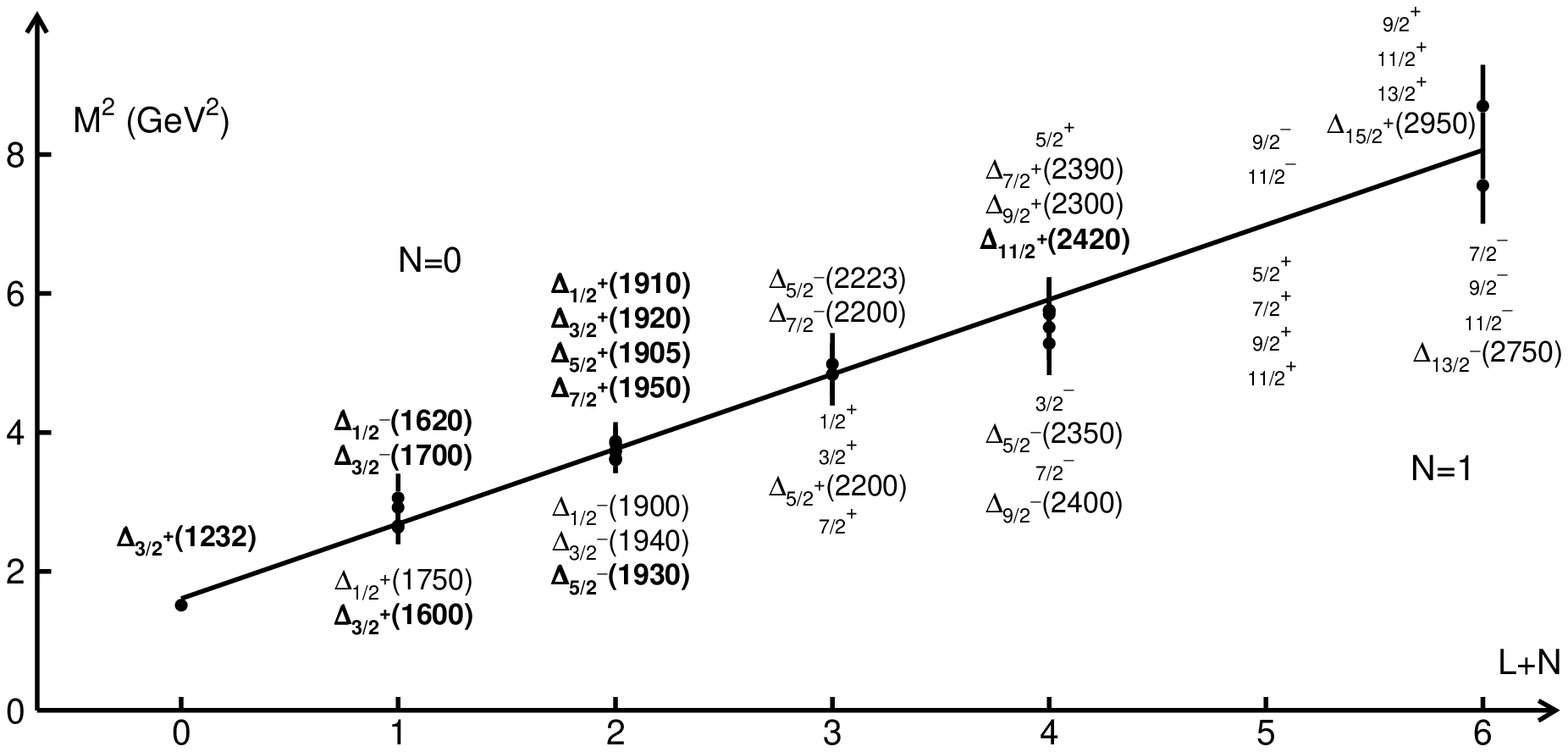}
\figcaption{\label{fig1}Masses of positive and negative parity
$\Delta$ resonances as a function of $L$+\textsf N. The masses are
three- and four-star resonances are bold, the others are classified
as one-star or two-star resonances. The so-called
$\Delta_{5/2^+}(2000)$ has entries at 1750\,MeV and at 2200\,MeV. We
retain the 2200\,MeV entry only.}
\end{center}
\begin{multicols}{2}
\noindent nances with even angular momenta, we assigned the quantum
numbers ($L$, $N=0$, $S=3/2$) or ($L$, $N=1$, $S=3/2$). $\Delta$
resonances with odd angular momenta, quantum numbers ($L$, $N=0$,
$S=1/2$) or ($L$, $N=1$, $S=3/2$) are assigned, with a strict
correlation between $N$ and $S$.

\subsection{Masses of nucleon resonances}
Masses of nucleon resonances depend not only on $L$ and $N$ but also
on $S$: the mass of the $L=1,S=1/2$ doublet - $N_{1/2^-}(1535)$ and
$N_{3/2^-}(1520)$ - is smaller than that of the $L=1,S=3/2$ triplet
comprising  $N_{1/2^-}(1650)$, $N_{3/2^-}(1700)$, and
$N_{1/2^-}(1675)$. The triplet is mass-degenerate with the
negative-parity $\Delta$ doublet. The mass of the spin doublet
$N_{3/2^+}(1720)$, $N_{5/2^-}(1680)$ with $L=2,S=1/2$ is smaller
than that of the  quartet with $L=2,S=3/2$ which is formed by
$N_{1/2^+}(1880)$, $N_{3/2^+}(1900)$, $N_{5/2^+}(1870)$,
$N_{7/2^+}(1990)$. The latter quartet is mass-degenerate with the
positive-parity quartet of $\Delta$ states having the same $L$ and
$S$. More examples can be found. Nucleon resonances with intrinsic
spin 1/2 have a mass which is smaller than their $S=3/2$ partners.
We assign a reduction in mass to those baryons which have a scalar
isoscalar diquark, a {\it good diquark}, as part of their wave
function. $\Delta$ resonances never have {\it good diquarks}, nor
nucleons with $S=3/2$. The nucleon has a wave function for which the
probability to find a good diquark $\alpha_D$ is equal to $1/2$. For
the two states $N_{1/2^-}(1535)$, $N_{3/2^-}(1520)$, $\alpha_D=1/4$,
and the squared mass difference to the spin or isospin 3/2-states is
half the $\Delta$--$N$ mass square difference. These observations
can be condensed into a surprisingly simple formula given by Forkel
and Klempt\cite{Forkel:2008un}
\begin{equation}
M^2 = a\cdot ({L}+{N}+3/2)-\,b\cdot \alpha_D\,\left[{\mathrm
GeV^2}\right]
\label{eq1}
\end{equation}
with $a=1.04$\,GeV$^2$ and $b=1.46$\,GeV$^2$. Eq. (\ref{eq1})
reproduces very well the full light-quark baryon mass spectrum.

\subsection{Other approaches}
It is instructive to compare the precision with which the different
models reproduce the baryon mass spectrum (Table \ref{tab3}). All
resonances from PDG\protect\cite{Amsler:2008zzb} are listed, 1-star
to 4-star but for resonances which are observed neither by
Arndt\protect\cite{Arndt:2006bf}, nor by
H\"ohler\protect\cite{Hohler:1979yr} nor by
Cutkovsky\protect\cite{Cutkosky:1980rh}, no mass is given here. Four
new states, suggested by BnGa and GWU analyses, are included.
Predictions based on AdS/QCD, on the quark model of Capstick-Isgur
model\cite{Capstick:1986bm} and on two variants of the Bonn
model\cite{Loring:2001kx} - differing in the choice of the Lorentz
\bc \tabcaption{\label{tab3}Masses of
$N$ and $\Delta$ resonances, experiment versus calculated masses.
Mass values and errors are taken from a recent review
\protect\cite{Klempt:2009pi}. FK: AdS/QCD model, eq.~(\ref{eq1}),
CI: Capstick and Isgur\protect\cite{Capstick:1986bm}, BnA and BnB
Bonn model\protect\cite{Loring:2001kx}, MK:  Skyrme model of
Karliner and Mattis\protect\cite{Karliner:1986wq}.
 }

\renewcommand{\arraystretch}{0.92}
\footnotesize
\begin{tabular}{ccccccc}
\toprule
Resonance\hspace{-3.mm}&\hspace{-3.mm}Mass\hspace{-3.mm}&\hspace{-3.mm} FK\hspace{-3.mm}&\hspace{-3.mm} CI \hspace{-3.mm}&\hspace{-3.mm}Bn-{\cal A}\hspace{-3.mm}&\hspace{-3.mm}Bn-{\cal B}\hspace{-3.mm}&\hspace{-3.mm} MK\\
$N_{\rm param}$\hspace{-3.mm}&\hspace{-3.mm}\hspace{-3.mm}&\hspace{-3.mm} 2 \hspace{-3.mm}&\hspace{-3.mm} 7 \hspace{-3.mm}&\hspace{-3.mm} 5\hspace{-3.mm}&\hspace{-3.mm}5\hspace{-3.mm}&\hspace{-3.mm}2 \\
\hline $N(940)$ \hspace{-3.mm}&\hspace{-3.mm}940\hspace{-3.mm}&\hspace{-3.mm}943 \hspace{-3.mm}&\hspace{-3.mm}960\hspace{-3.mm}&\hspace{-3.mm}939\hspace{-3.mm}&\hspace{-3.mm}939\hspace{-3.mm}&\hspace{-3.mm} 1190\\
$\Delta(1232)$\hspace{-3.mm}&\hspace{-3.mm}1232\;$\pm$\;1\hspace{-3.mm}&\hspace{-3.mm}1261\hspace{-3.mm}&\hspace{-3.mm}1230\hspace{-3.mm}&\hspace{-3.mm}1231\hspace{-3.mm}&\hspace{-3.mm}1261\hspace{-3.mm}&\hspace{-3.mm}1435\\

$N_{1/2^+}(1440)$\hspace{-3.mm}&\hspace{-3.mm}1450$\pm$32\hspace{-3.mm}&\hspace{-3.mm} 1396\hspace{-3.mm}&\hspace{-3.mm}1540 \hspace{-3.mm}&\hspace{-3.mm}1698 \hspace{-3.mm}&\hspace{-3.mm}1540\hspace{-3.mm}&\hspace{-3.mm}\\

$N_{1/2^-}(1535)$\hspace{-3.mm}&\hspace{-3.mm}1538$\pm$10\hspace{-3.mm}&\hspace{-3.mm} 1516\hspace{-3.mm}&\hspace{-3.mm} 1460\hspace{-3.mm}&\hspace{-3.mm}1435\hspace{-3.mm}&\hspace{-3.mm}1470\hspace{-3.mm}&\hspace{-3.mm}1478\\

$N_{3/2^-}(1520)$\hspace{-3.mm}&\hspace{-3.mm}1522\,$\pm$\,4\hspace{-3.mm}&\hspace{-3.mm} 1516 \hspace{-3.mm}&\hspace{-3.mm}1495\hspace{-3.mm}&\hspace{-3.mm}1476\hspace{-3.mm}&\hspace{-3.mm}1485\hspace{-3.mm}&\hspace{-3.mm}1715\\

$N_{1/2^-}(1650)$\hspace{-3.mm}&\hspace{-3.mm}1660$\pm$18\hspace{-3.mm}&\hspace{-3.mm} 1628 \hspace{-3.mm}&\hspace{-3.mm}1535\hspace{-3.mm}&\hspace{-3.mm}1660\hspace{-3.mm}&\hspace{-3.mm}1767\hspace{-3.mm}&\hspace{-3.mm}\\

$N_{3/2^-}(1700)$\hspace{-3.mm}&\hspace{-3.mm}1725$\pm$50\hspace{-3.mm}&\hspace{-3.mm} 1628 \hspace{-3.mm}&\hspace{-3.mm}1625\hspace{-3.mm}&\hspace{-3.mm}1606\hspace{-3.mm}&\hspace{-3.mm}1631\hspace{-3.mm}&\hspace{-3.mm}\\

$N_{5/2^-}(1675)$\hspace{-3.mm}&\hspace{-3.mm}1675\,$\pm$\,5\hspace{-3.mm}&\hspace{-3.mm} 1628 \hspace{-3.mm}&\hspace{-3.mm}1630\hspace{-3.mm}&\hspace{-3.mm}1655\hspace{-3.mm}&\hspace{-3.mm}1622\hspace{-3.mm}&\hspace{-3.mm}1744\\

$\Delta_{1/2^-}(1620)$\hspace{-3.mm}&\hspace{-3.mm}1626$\pm$23\hspace{-3.mm}&\hspace{-3.mm} 1628\hspace{-3.mm}&\hspace{-3.mm}1555\hspace{-3.mm}&\hspace{-3.mm}1654\hspace{-3.mm}&\hspace{-3.mm}1625 \hspace{-3.mm}&\hspace{-3.mm}1478\\

$\Delta_{3/2^-}(1700)$\hspace{-3.mm}&\hspace{-3.mm}1720$\pm$50\hspace{-3.mm}&\hspace{-3.mm} 1628 \hspace{-3.mm}&\hspace{-3.mm}1620\hspace{-3.mm}&\hspace{-3.mm}1628\hspace{-3.mm}&\hspace{-3.mm}1633\hspace{-3.mm}&\hspace{-3.mm}1737\\

$\Delta_{3/2^+}(1600)$\hspace{-3.mm}&\hspace{-3.mm}1615$\pm$80\hspace{-3.mm}&\hspace{-3.mm} 1628 \hspace{-3.mm}&\hspace{-3.mm}1795\hspace{-3.mm}&\hspace{-3.mm}1810\hspace{-3.mm}&\hspace{-3.mm}1923\hspace{-3.mm}&\hspace{-3.mm}1435\\

$N_{3/2^+}(1720)$ \hspace{-3.mm}&\hspace{-3.mm}1730$\pm$30\hspace{-3.mm}&\hspace{-3.mm} 1735 \hspace{-3.mm}&\hspace{-3.mm}1795\hspace{-3.mm}&\hspace{-3.mm}1688\hspace{-3.mm}&\hspace{-3.mm}1762\hspace{-3.mm}&\hspace{-3.mm}1982\\

$N_{5/2^+}(1680)$ \hspace{-3.mm}&\hspace{-3.mm}1683\,$\pm$\,3\hspace{-3.mm}&\hspace{-3.mm} 1735 \hspace{-3.mm}&\hspace{-3.mm}1770\hspace{-3.mm}&\hspace{-3.mm}1723\hspace{-3.mm}&\hspace{-3.mm}1718\hspace{-3.mm}&\hspace{-3.mm}1823\\

$N_{1/2^+}(1710)$ \hspace{-3.mm}&\hspace{-3.mm}1713$\pm$12\hspace{-3.mm}&\hspace{-3.mm} 1735 \hspace{-3.mm}&\hspace{-3.mm}1770\hspace{-3.mm}&\hspace{-3.mm}1729\hspace{-3.mm}&\hspace{-3.mm}1778\hspace{-3.mm}&\hspace{-3.mm}1427\\

$\Delta_{1/2^+}(1750)$\hspace{-3.mm}&\hspace{-3.mm}\hspace{-3.mm}&\hspace{-3.mm} - \hspace{-3.mm}&\hspace{-3.mm}1835\hspace{-3.mm}&\hspace{-3.mm}1866\hspace{-3.mm}&\hspace{-3.mm}1901\hspace{-3.mm}&\hspace{-3.mm}\\

$N_{1/2^-}(1905)$\hspace{-3.mm}&\hspace{-3.mm}1905$\pm$50\hspace{-3.mm}&\hspace{-3.mm} 1833 \hspace{-3.mm}&\hspace{-3.mm}1945\hspace{-3.mm}&\hspace{-3.mm}1910\hspace{-3.mm}&\hspace{-3.mm}1971\hspace{-3.mm}&\hspace{-3.mm}\\

$N_{3/2^-}(1860)$\hspace{-3.mm}&\hspace{-3.mm}1850$\pm$40\hspace{-3.mm}&\hspace{-3.mm} 1833 \hspace{-3.mm}&\hspace{-3.mm}1960\hspace{-3.mm}&\hspace{-3.mm}1940\hspace{-3.mm}&\hspace{-3.mm}1949\hspace{-3.mm}&\hspace{-3.mm}\\

\hspace{1.5mm}$N_{1/2^+}(1880)^a$\hspace{-3.mm}&\hspace{-3.mm}1890$\pm$50\hspace{-3.mm}&\hspace{-3.mm} 1926 \hspace{-3.mm}&\hspace{-3.mm}1880\hspace{-3.mm}&\hspace{-3.mm}1973\hspace{-3.mm}&\hspace{-3.mm}1974\hspace{-3.mm}&\hspace{-3.mm}\\

$N_{3/2^+}(1900)$\hspace{-3.mm}&\hspace{-3.mm}1940$\pm$50\hspace{-3.mm}&\hspace{-3.mm} 1926 \hspace{-3.mm}&\hspace{-3.mm}1870\hspace{-3.mm}&\hspace{-3.mm}1899\hspace{-3.mm}&\hspace{-3.mm}1904\hspace{-3.mm}&\hspace{-3.mm}\\

\hspace{1.5mm}$N_{5/2^+}(1870)^a$\hspace{-3.mm}&\hspace{-3.mm}1870$\pm$40\hspace{-3.mm}&\hspace{-3.mm} 1926 \hspace{-3.mm}&\hspace{-3.mm}1770\hspace{-3.mm}&\hspace{-3.mm}1934\hspace{-3.mm}&\hspace{-3.mm}1943\hspace{-3.mm}&\hspace{-3.mm}\\

$N_{7/2^+}(1990)$\hspace{-3.mm}&\hspace{-3.mm}2020$\pm$60\hspace{-3.mm}&\hspace{-3.mm} 1926 \hspace{-3.mm}&\hspace{-3.mm}2000\hspace{-3.mm}&\hspace{-3.mm}1989\hspace{-3.mm}&\hspace{-3.mm}1941\hspace{-3.mm}&\hspace{-3.mm}2011\\

$\Delta_{1/2^-}(1900)$\hspace{-3.mm}&\hspace{-3.mm}1910$\pm$50\hspace{-3.mm}&\hspace{-3.mm} 1926 \hspace{-3.mm}&\hspace{-3.mm}2035\hspace{-3.mm}&\hspace{-3.mm}2100\hspace{-3.mm}&\hspace{-3.mm}2169\hspace{-3.mm}&\hspace{-3.mm}2035\\

$\Delta_{3/2^-}(1940)$\hspace{-3.mm}&\hspace{-3.mm}1995$\pm$60\hspace{-3.mm}&\hspace{-3.mm} 1926  \hspace{-3.mm}&\hspace{-3.mm}2080\hspace{-3.mm}&\hspace{-3.mm}2122\hspace{-3.mm}&\hspace{-3.mm}2161\hspace{-3.mm}&\hspace{-3.mm}   \\

$\Delta_{5/2^-}(1930)$\hspace{-3.mm}&\hspace{-3.mm}1930$\pm$30\hspace{-3.mm}&\hspace{-3.mm} 1926  \hspace{-3.mm}&\hspace{-3.mm}2155\hspace{-3.mm}&\hspace{-3.mm}2170\hspace{-3.mm}&\hspace{-3.mm}2152\hspace{-3.mm}&\hspace{-3.mm} 1730  \\

$\Delta_{1/2^+}(1910)$\hspace{-3.mm}&\hspace{-3.mm}1935$\pm$90\hspace{-3.mm}&\hspace{-3.mm} 1926 \hspace{-3.mm}&\hspace{-3.mm}1835\hspace{-3.mm}&\hspace{-3.mm}1906\hspace{-3.mm}&\hspace{-3.mm}1928\hspace{-3.mm}&\hspace{-3.mm}1982\\

$\Delta_{3/2^+}(1920)$\hspace{-3.mm}&\hspace{-3.mm}1950$\pm$70\hspace{-3.mm}&\hspace{-3.mm} 1926 \hspace{-3.mm}&\hspace{-3.mm}1915\hspace{-3.mm}&\hspace{-3.mm}1910\hspace{-3.mm}&\hspace{-3.mm}1955\hspace{-3.mm}&\hspace{-3.mm}1946\\

$\Delta_{5/2^+}(1905)$\hspace{-3.mm}&\hspace{-3.mm}1885$\pm$25\hspace{-3.mm}&\hspace{-3.mm} 1926 \hspace{-3.mm}&\hspace{-3.mm}1910\hspace{-3.mm}&\hspace{-3.mm}1940\hspace{-3.mm}&\hspace{-3.mm}1932\hspace{-3.mm}&\hspace{-3.mm}1831\\

$\Delta_{7/2^+}(1950)$\hspace{-3.mm}&\hspace{-3.mm}1930$\pm$16\hspace{-3.mm}&\hspace{-3.mm} 1926 \hspace{-3.mm}&\hspace{-3.mm}1940\hspace{-3.mm}&\hspace{-3.mm}1956\hspace{-3.mm}&\hspace{-3.mm}1912\hspace{-3.mm}&\hspace{-3.mm}1816\\

$N_{1/2^+}(2100)$\hspace{-3.mm}&\hspace{-3.mm}2090$\pm$100\hspace{-3.mm}&\hspace{-3.mm} 2017 \hspace{-3.mm}&\hspace{-3.mm}1975\hspace{-3.mm}&\hspace{-3.mm}2127\hspace{-3.mm}&\hspace{-3.mm}2177\hspace{-3.mm}&\hspace{-3.mm}\\

$N_{1/2^-}(2090)$\hspace{-3.mm}&\hspace{-3.mm}\hspace{-3.mm}&\hspace{-3.mm} 2102 \hspace{-3.mm}&\hspace{-3.mm}2135\hspace{-3.mm}&\hspace{-3.mm}2200\hspace{-3.mm}&\hspace{-3.mm}2180\hspace{-3.mm}&\hspace{-3.mm} \\

$N_{3/2^-}(2080)$\hspace{-3.mm}&\hspace{-3.mm}2100$\pm$55\hspace{-3.mm}&\hspace{-3.mm} 2102 \hspace{-3.mm}&\hspace{-3.mm}2125\hspace{-3.mm}&\hspace{-3.mm}2079\hspace{-3.mm}&\hspace{-3.mm}2095\hspace{-3.mm}&\hspace{-3.mm} \\

$N_{5/2^-}(2060)^{a}$\hspace{-3.mm}&\hspace{-3.mm}2065$\pm$25\hspace{-3.mm}&\hspace{-3.mm} 2102 \hspace{-3.mm}&\hspace{-3.mm}2155\hspace{-3.mm}&\hspace{-3.mm}1970\hspace{-3.mm}&\hspace{-3.mm}2026\hspace{-3.mm}&\hspace{-3.mm} \\

$N_{7/2^-}(2190)$\hspace{-3.mm}&\hspace{-3.mm}2150$\pm$30\hspace{-3.mm}&\hspace{-3.mm} 2102 \hspace{-3.mm}&\hspace{-3.mm}2090\hspace{-3.mm}&\hspace{-3.mm} 2093\hspace{-3.mm}&\hspace{-3.mm}2100\hspace{-3.mm}&\hspace{-3.mm}2075\\

$N_{5/2^-}(2200)$\hspace{-3.mm}&\hspace{-3.mm}2160$\pm$85\hspace{-3.mm}&\hspace{-3.mm} 2102 \hspace{-3.mm}&\hspace{-3.mm}2234\hspace{-3.mm}&\hspace{-3.mm}2185\hspace{-3.mm}&\hspace{-3.mm}2217\hspace{-3.mm}&\hspace{-3.mm} \\

$N_{9/2^-}(2250)$\hspace{-3.mm}&\hspace{-3.mm}2255$\pm$55\hspace{-3.mm}&\hspace{-3.mm} 2184 \hspace{-3.mm}&\hspace{-3.mm} 2234\hspace{-3.mm}&\hspace{-3.mm}2212\hspace{-3.mm}&\hspace{-3.mm}2170\hspace{-3.mm}&\hspace{-3.mm}2234\\

$\Delta_{1/2^-}(2150)$\hspace{-3.mm}&\hspace{-3.mm}\hspace{-3.mm}&\hspace{-3.mm} 2184 \hspace{-3.mm}&\hspace{-3.mm}2140\hspace{-3.mm}&\hspace{-3.mm}2171\hspace{-3.mm}&\hspace{-3.mm}2217\hspace{-3.mm}&\hspace{-3.mm}\\

\hspace{1.5mm}$\Delta_{5/2^-}(2223)^b$\hspace{-3.mm}&\hspace{-3.mm}2223\,$\pm$\,53\hspace{-3.mm}&\hspace{-3.mm} 2184 \hspace{-3.mm}&\hspace{-3.mm}2155\hspace{-3.mm}&\hspace{-3.mm}2170\hspace{-3.mm}&\hspace{-3.mm}2179 \\

$\Delta_{7/2^-}(2200)$\hspace{-3.mm}&\hspace{-3.mm}2230\,$\pm$\,50\hspace{-3.mm}&\hspace{-3.mm} 2184 \hspace{-3.mm}&\hspace{-3.mm}2090\hspace{-3.mm}&\hspace{-3.mm}2210\hspace{-3.mm}&\hspace{-3.mm}2200\hspace{-3.mm}&\hspace{-3.mm} 2162\\

$N_{9/2^+}(2220)$\hspace{-3.mm}&\hspace{-3.mm}2360$\pm$125\hspace{-3.mm}&\hspace{-3.mm}2265\hspace{-3.mm}&\hspace{-3.mm}2327\hspace{-3.mm}&\hspace{-3.mm}2221\hspace{-3.mm}&\hspace{-3.mm}2221\hspace{-3.mm}&\hspace{-3.mm}2327\\

$\Delta_{7/2^+}(2390)$\hspace{-3.mm}&\hspace{-3.mm}2390$\pm$100\hspace{-3.mm}&\hspace{-3.mm} 2415 \hspace{-3.mm}&\hspace{-3.mm}2032\hspace{-3.mm}&\hspace{-3.mm}2340\hspace{-3.mm}&\hspace{-3.mm}2343\hspace{-3.mm}&\hspace{-3.mm}  \\

$\Delta_{9/2^+}(2300)$\hspace{-3.mm}&\hspace{-3.mm}2360$\pm$125\hspace{-3.mm}&\hspace{-3.mm}2415\hspace{-3.mm}&\hspace{-3.mm} 2407\hspace{-3.mm}&\hspace{-3.mm} 2453\hspace{-3.mm}&\hspace{-3.mm}2421\hspace{-3.mm}&\hspace{-3.mm}2407  \\

\hspace{1mm}$\Delta_{11/2^+}(2420)$\hspace{-3.mm}&\hspace{-3.mm}2462$\pm$120\hspace{-3.mm}&\hspace{-3.mm}2415\hspace{-3.mm}&\hspace{-3.mm}2450\hspace{-3.mm}&\hspace{-3.mm}2442\hspace{-3.mm}&\hspace{-3.mm}2388\hspace{-3.mm}&\hspace{-3.mm} 2327  \\

$\Delta_{9/2^-}(2400)$\hspace{-3.mm}&\hspace{-3.mm}2400$\pm$190\hspace{-3.mm}&\hspace{-3.mm} 2415 \hspace{-3.mm}&\hspace{-3.mm}2083\hspace{-3.mm}&\hspace{-3.mm}2280\hspace{-3.mm}&\hspace{-3.mm}2207\hspace{-3.mm}&\hspace{-3.mm} \\

$\Delta_{3/2^-}(2350)$\hspace{-3.mm}&\hspace{-3.mm}2310\,$\pm$\,85\hspace{-3.mm}&\hspace{-3.mm}2415\hspace{-3.mm}&\hspace{-3.mm}2145\hspace{-3.mm}&\hspace{-3.mm}2216\hspace{-3.mm}&\hspace{-3.mm}2234\hspace{-3.mm}&\hspace{-3.mm}\\

$N_{11/2^-}(2600)$\hspace{-3.mm}&\hspace{-3.mm}2630$\pm$120\hspace{-3.mm}&\hspace{-3.mm}2557\hspace{-3.mm}&\hspace{-3.mm}2327\hspace{-3.mm}&\hspace{-3.mm}2628\hspace{-3.mm}&\hspace{-3.mm}2610\hspace{-3.mm}&\hspace{-3.mm}2558\\

$N_{13/2^+}(2800)$\hspace{-3.mm}&\hspace{-3.mm}2800$\pm$160\hspace{-3.mm}&\hspace{-3.mm}2693\hspace{-3.mm}&\hspace{-3.mm}2558\hspace{-3.mm}&\hspace{-3.mm}2616\hspace{-3.mm}&\hspace{-3.mm}2619\hspace{-3.mm}&\hspace{-3.mm}2882\\

\hspace{1mm}$\Delta_{13/2^-}(2750)$\hspace{-3.mm}&\hspace{-3.mm}2720$\pm$100\hspace{-3.mm}&\hspace{-3.mm}2820\hspace{-3.mm}&\hspace{-3.mm}\hspace{-3.mm}&\hspace{-3.mm}2685\hspace{-3.mm}&\hspace{-3.mm}2604\hspace{-3.mm}&\hspace{-3.mm} 2579  \\

\hspace{1mm}$\Delta_{15/2^+}(2950)$\hspace{-3.mm}&\hspace{-3.mm}2920$\pm$100\hspace{-3.mm}&\hspace{-3.mm}2820\hspace{-3.mm}&\hspace{-3.mm}\hspace{-3.mm}&\hspace{-3.mm}2824\hspace{-3.mm}&\hspace{-3.mm}2768\hspace{-3.mm}&\hspace{-3.mm} 2810  \\
\bottomrule
\end{tabular}
\renewcommand{\arraystretch}{1.0}

$^{a}$: BnGa; $^{b}$:  GWU
\end{center}
structure of the confinement potential - are listed. For quark
models, the comparison is not fully straightforward, due to the
multitude of predicted states. The Bonn model\cite{Loring:2001kx}
predicts, e.g., for the $1/2^-$ sector two low mass states which are
readily identified, and then seven further states with masses, which
are found to be (1901, 1918); (2153, 2185, 2194, 2232, 2242) in
model $\cal A$, and (1971); (2082, 2180, 2203, 2261, 2270, 2345) in
model $\cal B$. We compare the experimental masses with the center
of gravity of a group of states. The groups were suggested by the
authors. At the conference there was the claim that an equally good
description of the data was obtained in a Skyrme
model\protect\cite{Karliner:1986wq}, also with just two parameters.
This claim is tested as well. From Table~\ref{tab3} we determine the
mean relative difference between calculated and measured mass for
the five
models:\\[-2ex]

\begin{tabular}{lccclccc}
\hspace{-7mm}$(\delta M/M)_{\rm
FK}$&\hspace{-3mm}=\hspace{-3mm}&\hspace{-3mm}2.5\%&\hspace{-2.5mm}(2p);\hspace{-2.5mm}&
\hspace{-2.5mm}$(\delta M/M)_{\rm
CI}$&\hspace{-3mm}=\hspace{-3mm}&\hspace{-3mm}5.6\%&\hspace{-2.5mm}(9p)\\
\hspace{-7mm}$(\delta M/M)_{\rm
BnA}$&\hspace{-3mm}=\hspace{-3mm}&\hspace{-3mm}5.1\%&\hspace{-2.5mm}(7p);
&\hspace{-2.5mm}$(\delta M/M)_{\rm
BnB}$&\hspace{-3mm}=\hspace{-3mm}&\hspace{-3mm}5.4\%&\hspace{-2.5mm}(7p)\\
\hspace{-7mm}$(\delta M/M)_{\rm
MK}$&\hspace{-3mm}=\hspace{-3mm}&\hspace{-3mm}9.1\%&\hspace{-2.5mm}(2p).
\end{tabular}\\

\noindent The number of parameters adjusted to achieve good
agreement with data is given in parentheses. At 2\,GeV mass, AdS/QCD
agrees on average within 50\,MeV, the quark models to about
110\,MeV, and the Skyrme model to about 190\,MeV. Compared to the
quark models, AdS/QCD requires substantially fewer parameters. The
Skyrme models fails to predict a large number of resonances,
including some well-established resonances, and gives the worst
description of the experimental mass spectrum.

\subsection{Interpretation}

Why is the mass formula derived from AdS/QCD - and suggested on a
phenomenological basis a few years earlier \cite{Klempt:2002vp} - so
successful? Two aspects are remarkable. First, in AdS/QCD the
coefficient $a$ is related to the hadron size, and the reduction in
mass of nucleons with good-diquark content is interpreted by a
smaller size of good diquarks compared to diquarks have spin or
isospin 3/2. Second, baryon resonances form super-multiplets with
defined $L$ and $S$. This is not the organization principle for the
dynamics of a highly relativistic three-quark system. Most physicist
prefer to stay with the highly relativistic three-quark system and
to abandon phenomenology. However, as mentioned in the introduction,
the nucleon mass is not understood as arising from the motion of
relativistic quarks but rather as effect of the breaking of chiral
symmetry of nearly massless quarks. Possibly, chiral symmetry
breaking is also the primary source for the masses of excited
baryons, but chiral symmetry is broken in an extended volume. A
physical picture emerges which assigns the largest fraction of the
masses of light-quark baryons to a volume in which field energy is
stored. Centrifugal forces expand the size as suggested a long time
ago by Nambu\cite{Nambu:1961tp}. The string-like behavior is the
reason why AdS/QFT works so nicely. Isoscalar scalar diquarks are
more tightly bound, their volume is smaller. The fraction of the
isoscalar scalar diquarks is smaller for odd angular momenta that in
case of even $L$. \vspace{4mm}

 \acknowledgments{I would like to
thank B. Metsch, H. Petry, and J.M. Richard for numerous clarifying
discussions, R. Beck for a critical reading of the manu\-script, and
all members of SFB/TR16 for continuous encouragement. Financial
support from the Deutsche Forschungsgemeinschaft (DFG) within the
SFB/TR16 is kindly acknowledged. }

\end{multicols}


\vspace{4mm}




\begin{thebibliography}{90}
\bibitem{Chew:1955zz}
  G.~F.~Chew and F.~E.~Low,
  Phys.\ Rev.\  {\bf 101}, 1570 (1956).
\bibitem{Kaiser:1995eg}
  N.~Kaiser, P.~B.~Siegel and W.~Weise,
  Nucl.\ Phys.\  A {\bf 594}, 325 (1995).
\bibitem{Jido:2003cb}
  D.~Jido, J.~A.~Oller, E.~Oset, A.~Ramos and U.~G.~Meissner,
  Nucl.\ Phys.\  A {\bf 725}, 181 (2003).
\bibitem{Glozman:this}L. Glozman,
``Chiral symmetry restoration in excited hadrons and dense matter",
this conference.
\bibitem{Amsler:2008zzb}
  C.~Amsler {\it et al.},
  Phys.\ Lett.\  B {\bf 667}, 1 (2008).
\bibitem{Arndt:2006bf}
  R.~A.~Arndt, W.~J.~Briscoe, I.~I.~Strakovsky and R.~L.~Workman,
  Phys.\ Rev.\  C {\bf 74}, 045205 (2006)\bibitem{Hohler:1979yr}
  G.~H\"ohler, F.~Kaiser, R.~Koch and E.~Pietarinen,
  ``Handbook Of Pion Nucleon Scattering,''
 Fachinform. Zentr. Karlsruhe 1979, 440 P. (Physics Data, No.12-1 (1979)).
\bibitem{Cutkosky:1980rh}
  R.~E.~Cutkosky, C.~P.~Forsyth, J.~B.~Babcock, R.~L.~Kelly and R.~E.~Hendrick,
  ``Pion - Nucleon Partial Wave Analysis,''
4th Int. Conf. on Baryon Resonances, Toronto, Canada, Jul 14-16,
1980. Published in Baryon 1980:19 (QCD161:C45:1980).
\bibitem{Anisovich:2009pr}
 A.V.~Anisovich {\it et al.},
``Photoproduction of pions and properties of baryon resonances", in
preparation.
\bibitem{Brodsky:this}S.J. Brodsky, ``AdS/QCD and Light
Front Holography: A new approximation to QCD", this conference.
\bibitem{Karch:2006pv}
  A.~Karch, E.~Katz, D.~T.~Son and M.~A.~Stephanov,
  Phys.\ Rev.\  D {\bf 74}, 015005 (2006).
\bibitem{Forkel:2007cm}
  H.~Forkel, M.~Beyer and T.~Frederico,
  JHEP {\bf 0707}, 077 (2007).
\bibitem{Klempt:2007cp}
  E.~Klempt and A.~Zaitsev,
  Phys.\ Rept.\  {\bf 454}, 1 (2007).
\bibitem{Forkel:2008un}
  H.~Forkel and E.~Klempt,
  Phys.\ Lett.\  B {\bf 679}, 77 (2009).
\bibitem{Capstick:1986bm}
  S.~Capstick and N.~Isgur,
  Phys.\ Rev.\  D {\bf 34}, 2809 (1986).
\bibitem{Loring:2001kx}
U.~L\"oring {\it et al.}, 
Eur.\ Phys.\ J.\ A {\bf 10},  395, 447 (2001).
\bibitem{Karliner:1986wq}
  M.~Karliner and M.~P.~Mattis,
  Phys.\ Rev.\  D {\bf 34}, 1991 (1986).
\bibitem{Klempt:2009pi}
  E.~Klempt and J.~M.~Richard,
  ``Baryon spectroscopy,''
  arXiv:0901.2055 [hep-ph].
\bibitem{Klempt:2002vp}
  E.~Klempt,
  Phys.\ Rev.\  C {\bf 66}, 058201 (2002).
\bibitem{Nambu:1961tp}
  Y.~Nambu and G.~Jona-Lasinio,
  Phys.\ Rev.\  {\bf 122}, 345 (1961).

\end{thebibliography}
\end{document}